\documentclass{article}

\usepackage{amscd}
\usepackage{amsthm}
\usepackage{amsmath}
\usepackage{amssymb}
\usepackage{xypic}
\usepackage[latin5]{inputenc}
\usepackage[english]{babel}
\usepackage{graphicx}
\usepackage{float}

\newtheorem{thm}{Theorem}[section]

\theoremstyle{definition}
\newtheorem{defn}[thm]{Definition}

\theoremstyle{remark}

\numberwithin{equation}{subsection}

\begin{document}

\title{\textsc{A characterization of the electromagnetic stress-energy tensor}}
\author{Navarro, J.\thanks{The first author has been partially supported by Junta de Extremadura
and FEDER funds. \newline
{\it Email addresses:} navarrogarmendia@unex.es, jsancho@unex.es \newline
Department of Mathematics, University of Extremadura,
Avda. Elvas s/n, 06071, Badajoz, Spain}
\and Sancho, J.B. }

%\date{\today}
\date{November 2010}

\maketitle

\begin{abstract}
 In \cite{Einstein}, we pointed out how a dimensional analysis of the stress-energy tensor of the gravitational field allows to derive the field equation of General Relativity.

 In this note, we comment an analogous reasoning in presence of a 2-form, that allows to characterize the so called electromagnetic stress-energy tensor.
\end{abstract}

%%%%%%%%%%%%%%%%%%%%%%%%%%%%%%%%%%%%%%%%%%%%
%% MAINMATTER
%%%%%%%%%%%%%%%%%%%%%%%%%%%%%%%%%%%%%%%%%%%%

\subsection*{Natural tensors associated to a metric and a 2-form}

Let $X$ be an $n$ dimensional smooth manifold. Let us recall the rigorous definition of an ``intrinsic and local construction from a metric and a 2-form$"$ (for a more detailed account in the case of solely a metric, see \cite{Einstein} and references therein).

\medskip Let $S^2_+T^*X \to X$ be the fibre bundle of semiriemannian metrics of a given signature on $X$, $\Lambda^2 T^*X \to X$ be the vector bundle of 2-forms on $X$ and
$\bigotimes^p T^*X \otimes \bigotimes^q TX \to X$ be the vector bundle of $(p,q)$-tensors on $X$. Let us also denote their sheaves of smooth sections by \textit{Metrics}, \textit{Forms}
and \textit{Tensors}, respectively.

A morphism of sheaves $T \colon Metrics \times \textit{Forms} \to Tensors$ is said to be \textit{natural} if it is equivariant with respect to the action of local diffeomorphisms of $X$; that is, if for each diffeomorphism $\tau \colon U \to V$ between open sets of $X$ and for each metric $g$ and each 2-form $F$ on $V$, the
following condition is satisfied:
\begin{equation}\label{Naturalidad} T(\tau^* g , \tau^* F) = \tau^* (T(g , F))\ . \end{equation}

\begin{defn} A tensor \textbf{naturally constructed} from a metric $g$ and a 2-form $F$ (or a \textbf{natural tensor} associated to $(g , F)$) is a tensor of the type $T(g , F)$, where
$T \colon \textit{Metrics} \times \textit{Forms} \rightarrow \textit{Tensors}$ is a natural morphism of sheaves. \footnote{There is also a regularity hypothesis for $T$, that amounts to saying that the construction is ``differentiable$"$, see \cite{Einstein} or \cite{Kolar}.}
\end{defn}

   The main examples of tensors naturally constructed from a metric are the curvature tensor of the metric, its covariant derivatives and tensor products and contractions of these.

The main example of a 2-covariant tensor $E_2(g, F)$ naturally constructed from a metric $g$ and a 2-form $F$ is the so called electromagnetic stress-energy tensor, that in local coordinates is written as:
$$ E_{ij} = F_{i}^{\ k} F_{kj} - \frac{1}{4}\, F_{kl} F^{kl} g_{ij} $$

 Nevertheless notice that, by definition, the coefficients of a natural tensor $T(g , F)$ are only assumed to be locally constructed from $(g , F)$ and therefore need not necessarily be polynomial functions on the coefficients of $g$, $g^{-1}$, $F$ and its derivatives.

\paragraph{Dimensional analysis} In the theory of General Relativity, space-time is a smooth manifold $X$ of dimension 4 endowed with a Lorentz
metric $g$, called the time metric. The proper time of a
particle following a trajectory in $X$ is defined to be the length of that curve using the metric $g$. So that if the
metric $g$ is changed by a proportional one $\lambda^2 g$, with $\lambda \in \mathbb{R}^+$, then the proper time of
particles is multiplied by the factor $\lambda$. Therefore, replacing the metric $g$ by $\lambda^2g$ amounts to a
change in the time unit.

On the other hand, the presence of an electromagnetic field on the space-time $X$ is represented with a 2-form $F$. The endomorphism $\bar{F}$ metrically equivalent to $F$ assigns to each observer $u$ carrying the unit of charge, the acceleration that he measures due to the electromagnetic field, $\nabla_u u$ (recall the Lorentz equation). So, if we change the time unit and therefore replace the time metric $g$ by $\lambda^2 g$, then the observer $u$ becomes $\lambda^{-1} u$; the endomorphism $\bar{F}$ becomes $\lambda^{-1} \bar{F}$ and the 2-form $F$ changes to $\lambda F$.

% La ecuaci\'{o}n de Lorentz es m u^\nabla u  = q \bar{F} (u)
% y como tomamos todas las constantes universales iguales a 1, la masa y la carga cambian del mismo modo al cambiar la unidad
% de tiempo, y solo hay que estudiar lo demas.

\medskip
These reasonings lead to consider a special kind of homogeneity for the natural constructions:

\begin{defn} A morphism of sheaves $T \colon \textit{Metrics} \times \textit{Forms} \rightarrow \textit{Tensors}$ is
said to be homogenous of \textbf{weight $w \in \mathbb{R}$} if it satisfies:
$$ T(\lambda^2 g , \lambda F) = \lambda ^w \, T(g , F) \, \qquad \quad \forall \, g , F \ \    \forall \,
\lambda > 0 \ .$$ If the morphism $T$ has weight 0, it is said to be \textbf{independent of the unit of time}.

A tensor $T(g , F)$ naturally constructed from a metric and a 2-form is homogenous of \textbf{weight} $w$ if the corresponding morphism of
sheaves $T$ is homogenous of weight $w$.\end{defn}

\subsubsection*{Description of the vector space of natural tensors}

To describe the space of homogenous natural tensors $T(g, F)$ of weight $w$ (that results to be a finite dimensional $\mathbb{R}$-vector space), it is useful to introduce normal tensors at a point.

Fix any point $x\in X$. In the case of a metric, these vector spaces of normal tensors $N_r$ are described in \cite{Einstein}. In order to deal with a 2-form, we also introduce:
$$ \Lambda_l := \Lambda ^2 T^*_xX \otimes S^l T^*_xX $$ If $l=0$, $\Lambda_0$ is defined to be the space of 2-forms at $x\in X$.

On a semi-riemannian manifold, the germ on $x$ of a 2-form $F$ produces a sequence of tensors in $\Lambda_l$, $l \geq 0$. To see this, fix normal coordinates in a neighbourhood of $x$ and write:
$$ F^l_x := \sum_{ijk_1 \ldots k_l} F_{ij, k_1 \ldots k_l} \, dx_i \wedge dx_k \otimes dx_{k_1} \otimes \ldots \otimes dx_{k_l} \in \Lambda_l $$ where:
$$ F_{ij,k_1 \ldots k_l} := \frac{\partial^l F_{ij}}{\partial x_{k_1} \cdots \partial x_{k_l}} (x) $$

Moreover, let $O := O(n^+ , n^-)$ be the orthogonal group of $(T_{x}X, g_{x})$ and let $T_{p,x}^q $ be the space of $p$-covariant and $q$-contravariant tensors at $x$.

The vector space of homogenous natural tensors $T(g,F)$ is described in the following theorem, whose proof is completely analogous to the one referred in \cite{Einstein}:

\begin{thm}\label{Stredder} There exists an $\mathbb{R}$-linear isomorphism:
$$\begin{array}{c}
  \{ \mbox{(p,q)-Tensors of weight $w$ naturally constructed from $(g,F)$} \}  \\ \medskip
  \parallel \\ \medskip
   \bigoplus\limits_{\{ d_i , c_j \} } \mathrm{Hom}_{O} ( \ S^{d_2} N_2 \otimes \cdots \otimes
S^{d_r} N_r \otimes S^{c_0} \Lambda_0 \otimes \ldots \otimes S^{c_l} \Lambda_l \, \ ,\ \,  T_{p,x}^q \ )
\end{array}$$
where the summation is over all sequences of non-negative integers $\{d_2 , \ldots , d_r , c_0 , \ldots , c_l \} $, $r\geq 2$, satisfying
the equation:
\begin{equation}\label{Condicion}
2d_2 + \ldots + r\, d_r + c_0 + 2c_1 + \ldots + (s+1) c_s = p-q -w\ .
\end{equation}
If this equation has no solutions, the above vector space reduces to zero.
\end{thm}

The isomorphism of the above theorem is explicit (see Remarks in \cite{Einstein}) and can be used, in simple cases, to make exhaustive computations. As an application, it follows the announced characterization.

\subsection*{A characterization of the electromagnetic stress-energy tensor}

Let us consider the natural tensor $E_2(g,F)$ defined in local coordinates as:
$$ E_{ij} = F_{i}^{\ k} F_{kj} - \frac{1}{4}\, F_{kl} F^{kl} g_{ij} $$

There already exists a geometric characterization of this tensor (see \cite{Kerrighan}). The following one uses homogeneity (condition \textit{(a)} below) to get rid of the assumptions imposed in \cite{Kerrighan} on the derivatives of the metric and the 2-form that appear on  the tensor.

\begin{thm} The tensor $E_2$ is the only (up to constant multiples) 2-covariant tensor naturally constructed from a metric and a 2-form satisfying:
\begin{itemize}
\item[(a)] It is independent of the unit of time (i.e., $E_2 ( \lambda ^2 g , \lambda F ) = E_2 ( g, F)$, $\forall \lambda > 0$  ).

\item[(b)] If $dF = \delta F = 0$, then $\mathrm{div}\, E_2 = 0$.

\item[(c)] If $F = 0$, then $E_2 = 0$ (i.e., there are no addends independent of $F$).

\end{itemize}
\end{thm}

\proof Using Theorem \ref{Stredder}, we only have to analyze the solutions to:
$$ 2d_2 + \ldots + r\, d_r + c_0 + 2c_1 + \ldots + (s+1) c_s = 2 $$

Condition \textit{(c)} says that we cannot consider solutions where $c_0 = c_1 = \ldots c_s = 0$. So there are only two possibilities:
\begin{itemize}
\item $c_1 = 1, c_i = d_j = 0$. In this case we have to compute the $O$-invariant linear maps (see Remarks in \cite{Einstein}):
$$ \Lambda^2 T^*_x X \otimes T^*_x X \otimes T^*_x X \otimes T^*_x X \to \mathbb{R}$$
This space of maps is generated, as an $\mathbb{R}$-vector space, by successive contractions. As the tensor space has in this case an odd number of indices, there only exists the null map.

\item $c_0 = 2 , c_i = d_j = 0$. For this case we have to compute the $O$-invariant linear maps:
    $$ S^2 (\Lambda^2 T^*_x X ) \otimes T^*_xX \otimes T^*_xX \to \mathbb{R}$$
    Due to the symmetries, there are only two independent generators:
    $$ (13)(24)(56) \quad \mathrm{ and } \quad (13)(25)(46) $$ where $(ij)$ denotes contraction of that pair of indices. These maps correspond, respectively, to the tensors:
    $$ F_{kl} F^{kl} g_{ij} \quad \mathrm{ and } \quad F_i^{\ k} F_{kj} $$
\end{itemize}

Now, the divergence condition \textit{(b)} implies, by a standard computation, that the only possible natural
tensors are constant multiples of $E_2$. \hfill $\square$

\paragraph{Super-energy tensor of a $p$-form:} The above characterization can be readily generalized to the super-energy tensor of a $p$-form (see \cite{Senovilla}), although the dimensional analysis is no longer made a priori, as it has been done before in the case of a 2-form.

\begin{thm} Let $\omega_p$ be a $p$-form on a semi-riemannian manifold $(X, g)$. The super-energy tensor of the $p$-form $T_2 (g, \omega_p)$ is the only (up to constant multiples) 2-covariant tensor naturally constructed from $g$ and $\omega_p$ satisfying:
\begin{itemize}
\item  $T_2 ( \lambda ^2 g , \lambda^{p-1} \omega_p ) = T_2 ( g, \omega_p )$, $\forall \lambda \in \mathbb{R}$.

\item If $dF = \delta F = 0$, then $\mathrm{div}\, T_2 = 0$.

\item $F = 0$ if and only if $T_2 = 0$.

\end{itemize}
\end{thm}

%%%%%%%%%%%%%%%%%%%%%%%%%%%%%%%%%%%%%%%%%%%%%%%%
%% BACKMATTER
%%%%%%%%%%%%%%%%%%%%%%%%%%%%%%%%%%%%%%%%%%%%%%%%

%\begin{theacknowledgments}
%\end{theacknowledgments}


\begin{thebibliography}{9}


\bibitem{Einstein}
J. Navarro  and J.B.~Sancho, \emph{J. of Geom. and Phys.} \textbf{58},
  1007--1014 (2008).

\bibitem{Kerrighan} D.B. Kerrighan, \emph{J. of Math. Phys.}, \textbf{10}, (1982)

\bibitem{Kolar} I. Kol\'{a}r, P.W. Michor, J. Slov\'{a}k: \emph{Natural operations in differential geometry},
Springer-Verlag, Berlin, 1993, p 171.


\bibitem{Senovilla} J.M.M. Senovilla, \emph{Class. Quant. Grav.} \textbf{17}, 2799-2841, (2000)



\end{thebibliography}
\end{document}